\documentclass[a4paper]{article}
\usepackage{Odyssey2018}
\usepackage{epsfig,amssymb,amsmath,multirow,boldline,graphicx,makecell,mathtools,hyperref}
\ninept
\usepackage{kotex}
\usepackage{color,soul}

\title{ \textbf{{\LARGE MCE 2018}}:\\ {\Large The 1st Multi-target speaker detection and identification\\ Challenge Evaluation (MCE) Plan, Dataset and Baseline System}\thanks{Document Ver. July 17, 2018}}

\makeatletter
\def\name#1{\gdef\@name{#1\\}}
\makeatother
\name{Suwon Shon$^1$, Najim Dehak$^2$, Douglas Reynolds$^3$, James Glass$^1$}
\address{MIT Computer Science and Artificial Intelligence Laboratory,  Cambridge, MA, USA$^1$\\
Center for Language and Speech Processing, Johns Hopkins University, Baltimore, USA$^2$\\
MIT Lincoln Laboratory, Lexington, MA, USA$^3$ \\ \\ \texttt{\href{http://www.mce2018.org/}{http://www.mce2018.org/}} }

\begin{document}
\maketitle

\section{Introduction}

The Multitarget Challenge aims to assess how well current speech technology is able to determine whether or not a recorded utterance was spoken by one of a large number of ``blacklisted'' speakers.  It is a form of multi-target speaker detection based on real-world telephone conversations.  Data recordings are generated from call center customer-agent conversations.  Each conversation is represented by a single i-vector~\cite{Dehak2011}.  Given a pool of training and development data from non-Blacklist and Blacklist speakers, the task is to measure how accurately one can detect 1) whether a test recording is spoken by a Blacklist speaker, and 2) which specific Blacklist speaker was talking.
%All data used in our multitarget detection experiment are real-world data coming from actual customer-client conversations in a call center. We collected a group of blacklist speakers, all of whom are previously identified fraudsters. We try to measure how accurately we can detect whether a speaker is in the blacklist. We will also measure how accurately we can identify the specific speaker in the blacklist. 
Although the primary task will restrict participants to the provided data, participants are allowed to submit secondary systems that use additional data in order to achieve better performance. 

\section{Task Description : Multi-target detection and identification}
\subsection{Task definition}
The task of MCE2018 is \textit{multi-target (speaker) detection and identification}. At first, the task verifies a given speech segment is from blacklist cohort or not. Second, if the input segment turns out as a blacklist, then need to identify who it is among blacklist speaker set. Thus, the performance evaluation will be done using two different metrics: Top-S stack detector and Top-1 stack detector. S is blacklist speaker size, so top-S stack detector only detects the input speech is from blacklist cohort or not. Top-1 stack detector not only detects blacklist cohort but also verifying speaker identify among blacklist. 

The task of MCE 2018 is similar to speaker verification task which verifies given speech segment and the target speaker enrollment data are from the same speaker or not. However, on this task, we only detect the identity if the input speech is spoken by the blacklist.

\subsection{Evaluation rules}
The participants are free to use the training and development set as they want. The test set should not be used for any purpose. Each register can submit up to six result files for evaluation and at least a single file is required for all participants as table~\ref{tab:condition}.

\noindent
\textbf{Fixed condition} : The fixed condition limits the system training to data provided from the MCE 2018 organizer.

\noindent
\textbf{Open condition} : The open condition does not have any limitation to use any dataset to training the system.

% \usepackage{multirow}
% \usepackage{graphicx}
\begin{table}[h]
\centering
\caption{Two condition and up to six submission files}
\label{tab:condition}
\resizebox{0.7\linewidth}{!}{%
\begin{tabular}{|c|c|c|}
\hline
\multirow{2}{*}{Submission} & \multicolumn{2}{c|}{Training condition} \\ \cline{2-3} 
 & Fixed & Open \\ \hline\hline
Primary & \textbf{Required} & Optional \\ \hline
Contrastive 1 & Optional & Optional \\ \hline
Contrastive 2 & Optional & Optional \\ \hline
\end{tabular}%
}
\end{table}

\section{Data Description}
Data will be provided in the form of i-vectors, which were extracted from the raw audio.  A standard train and development set will be provided to participants.  Each dataset consists of both blacklist (watchlist) speakers and non-blacklist (background) speakers. All 3,631 blacklist speakers appear 3 times in the train set and once in the development set, although the blacklist speaker labels between the train and development sets are different.  No information will be provided about the distribution of speakers in the test set, in order to better reflect the real-world scenario. 
% \jgcomment{we should not explain the test set distribution, as people can game the result if they know that each blacklist speaker appears once in the test set} 
%\swscomment{Is there any overlap on Background speakers? or they are complete different speaker?}

\subsection{Train Set}
Table~\ref{tab:train} shows the distribution of data for the training set.  Speaker labels are provided for this dataset.

\begin{table}[ht]
\centering
\resizebox{\linewidth}{!}{%
\begin{tabular}{|c|c|c|c|}
\hline
Subset     & \# of speakers & \# of utts. per speaker                     & Total utts. \\ \hline
Blacklist  & 3,631           & 3 & 10,893      \\ \hline
Background & 5,000           & $\ge$4 & 30,952      \\ \hline
\end{tabular}%
}
\caption{Training dataset description.}
\label{tab:train}
\end{table}

\subsection{Development Set}
Table~\ref{tab:dev} shows the distribution of speakers for the development set.  Speaker labels will be provided for the blacklist subset for performance validation.  Participants are free to use the development set for any purpose such as validation or training.
\begin{table}[ht]
\centering
\resizebox{\linewidth}{!}{%
\begin{tabular}{|c|c|c|c|}
\hline
Subset & \# of speakers & \# of utts. per speaker & Total utts. \\ \hline
Blacklist & 3,631 & 1 & 3,631\\ \hline
Background & 5,000& 1 & 5,000\\ \hline
\end{tabular}%
}
\caption{Development dataset description.}
\label{tab:dev}
\end{table}

\subsection{Test Set}
No label will be provided for the test dataset. All performance measurements will be made on the test set. Participants should {\it not} use the set for training or tuning of any kind purpose.

%\swscomment{should we hide this statistics? test set will be provided in random order of course.}
%\begin{table}[ht]
%\centering
%\resizebox{\linewidth}{!}{%
%\begin{tabular}{|c|c|c|c|}
%\hline
%Subset & \# of speakers & \# of utts. per speaker & Total utts. \\ \hline
%Blacklist & 3631 & 1 & 3631\\ \hline
%Background & 12386& 1 & 12386\\ \hline
%\end{tabular}%
%}
%\caption{Test dataset description}
%\label{tab:test}
%\end{table}

\subsection{Data format}
\vspace{2mm}

The i-vector and its speaker identification label will be provided in the following CSV format files:\\ 

\noindent 
\texttt{trn\_blacklist.csv}\\
\texttt{trn\_background.csv}\\
\texttt{dev\_blacklist.csv}\\
\texttt{dev\_background.csv}\\
\texttt{tst\_mix.csv}\\
\texttt{bl\_matching.csv}\\
\vspace{2mm} 

\noindent Each line in the file will contain one utterance id and 600 real numbers (i.e., the i-vector), separated by white space. The first four characters in the utterance id correspond to the speaker id. An example is shown below.\\

\texttt{aagj\_239446,1.1359440, ... ,-0.6017886}

\noindent \texttt{bl\_matching.csv} file contain unique blacklist speaker id and its corresponding speaker id in each dataset. An example is shown below

\texttt{50399530,dev\_fvth,train\_phee}

\noindent Both \texttt{fvth} in dev set and \texttt{phee} in train set is same speaker with unique id \texttt{50399530}

\section{Data Preparation}

A simple energy-based voice activity detector was used to extract speech frames from the original audio data.  Speech frames are represented by 60 dimensional MFCC vectors (i.e., 20 MFCCs + 20 delta MFCCs + 20 delta-delta MFCCs).  A 4,096 component Gaussian mixture model (GMM) is created from the training data and used as the universal background model (UBM), and used as the basis for i-vector extraction.  600-dimensional i-vectors~\cite{Dehak2011} are subsequently generated for the speech frames excised from each recording.\\

\noindent
Sampling Frequency: 8kHz\\
Frame duration: 20ms\\
MFCC dimension: 20\\
GMM-UBM components: 4,096\\
i-vector dimension: 600

\section{Performance Measures}

Performance will be reported using the equal error rate (EER) metric which is calculated in a similar fashion as conventional 1-1 speaker verification tasks. For a single target detector for a conventional speaker verification task, the miss and false alarm (FA) probability is given by
\begin{equation}
P_{Miss}(\theta) = P(y<\theta | C_x=C)
\end{equation}
\begin{equation}
P_{FA}(\theta) = P(y>\theta | C_x\neq C)
\end{equation}
where $\theta$ is an accept/reject decision threshold, $y$ is the similarity score for hypothesis $h$ that input test utterance $x$ of class $C_x$ belongs to class $C$. An acceptance is made if the score $y$ is above threshold $\theta$, and a rejection occurs when the score is below the threshold.  For a given decision threshold $\theta$, $P_{Miss}(\	theta)$ measures the fraction of incorrect rejections that are made when the hypothesized class $C$ corresponds to the true class $C_x$, while $P_{FA}(\theta)$ measures the fraction of accepts that are incorrectly made when hypothesized class $C$ does not correspond to the true class.

%The false acceptance rate (FAR) is equal to the percentage of non-target speakers who are falsely identified as target speakers:
%\begin{equation}
%FAR=P\bigg(\argmax_{y\in C} s(x,y) \geq \theta|x \not \in C\bigg),
%\end{equation}
%and the false rejection rate (FRR) is equal to the percentage of target speakers who are falsely identified as non-target speakers:
%\begin{equation}
%FRR=P\bigg(\argmax_{y\in C} s(x,y) <\theta|x \in C\bigg),
%\end{equation}

%where $s(x,y)$ is the score between utterances $x$ and $y$, $\theta$ is the decision threshold, and $C$ is the set of target speakers.

%To keep our results consistent for comparison, we adopt a TOP-1 approach for all experiments. That is, we only consider the highest normalized score in the blacklist. 

The basic $P_{Miss}$ and $P_{FA}$ are modified to create two metrics that will be used for this task: the Top-S detector, and the Top-1 detector~\cite{Singer2004}.  The Top-S detector must decide if a test vector belongs to {\it any} of the blacklist speakers or not.  The Top-1 detector must decide if a test vector corresponds to a {\it particular} blacklist speaker or not.  

\subsection{Top-S stack detector (Multi-target cohort detection)}
Given the total number of blacklist speakers, $S$, the Top-S stack detector determines if the test input belongs to any of the blacklist speakers.  The detector produces a set of scores, $y_1,...,y_S$ corresponding to the set of class hypotheses $h_1,...,h_S$.  The blacklist score $y^*$ corresponds to the maximum of all blacklist speaker scores $\{y_1,...,y_S\}$.
A miss occurs when is below the threshold ($y^*<\theta$) if the input is spoken by a blacklist speaker. Similarly, a false alarm occurs when the $y^*$ is above the accept threshold ($y>\theta$) when in fact the input is not from a blacklist speaker.

%$y^*=$Max$\{y_i\}$ where $i$ is set of blacklist class.

\begin{equation}
P_{Miss}(\theta) = P(y^*<\theta | C_x\in \{C_{1,...,S}\})
\end{equation}
\begin{equation}
P_{FA}(\theta) = P(y^*>\theta | C_x\not\in \{C_{1,...,S}\})
\end{equation}

Note that although $y^*$ is defined as a maximum of all blacklist speaker hypothesis scores, $y^*$ could be computed via some other function of the hypothesis scores.  For evaluation, all that is required is that each test input have a generated score, $y^*$.

\subsection{Top-1 stack detector (Multi-target identification) }
The Top-1 stack detector also detects blacklist speakers but determines if the test input is spoken by one particular blacklist speaker.  Thus, there is new type of error for this task which is a form of confusion error.  The confusion error means that an actual blacklist input is correctly detected as a blacklist speaker, but fails to correctly identify the speaker. The confusion error occurs if score $y^*$ is above threshold $\theta$, but $C_x$ does not correspond to the class hypothesis of $h^*$.

\begin{equation}
\begin{split}
P_{Miss}(\theta) = & P(y^*<\theta) | C_x\in \{C_{1,...,S}\}\}) \\
& + P(y^*>\theta,C_x \neq h^* | C_x\in \{C_{1,...,S}\})
\end{split}
\end{equation}

\begin{equation}
P_{FA}(\theta) = P(y^*>\theta | C_x\not\in \{C_{1,...,S}\})
\end{equation}
Note that $P_{FA}$ is the same for both metrics.

\section{Baseline}
\begin{figure}[h]
    
    \centering
    \includegraphics[width=\linewidth]{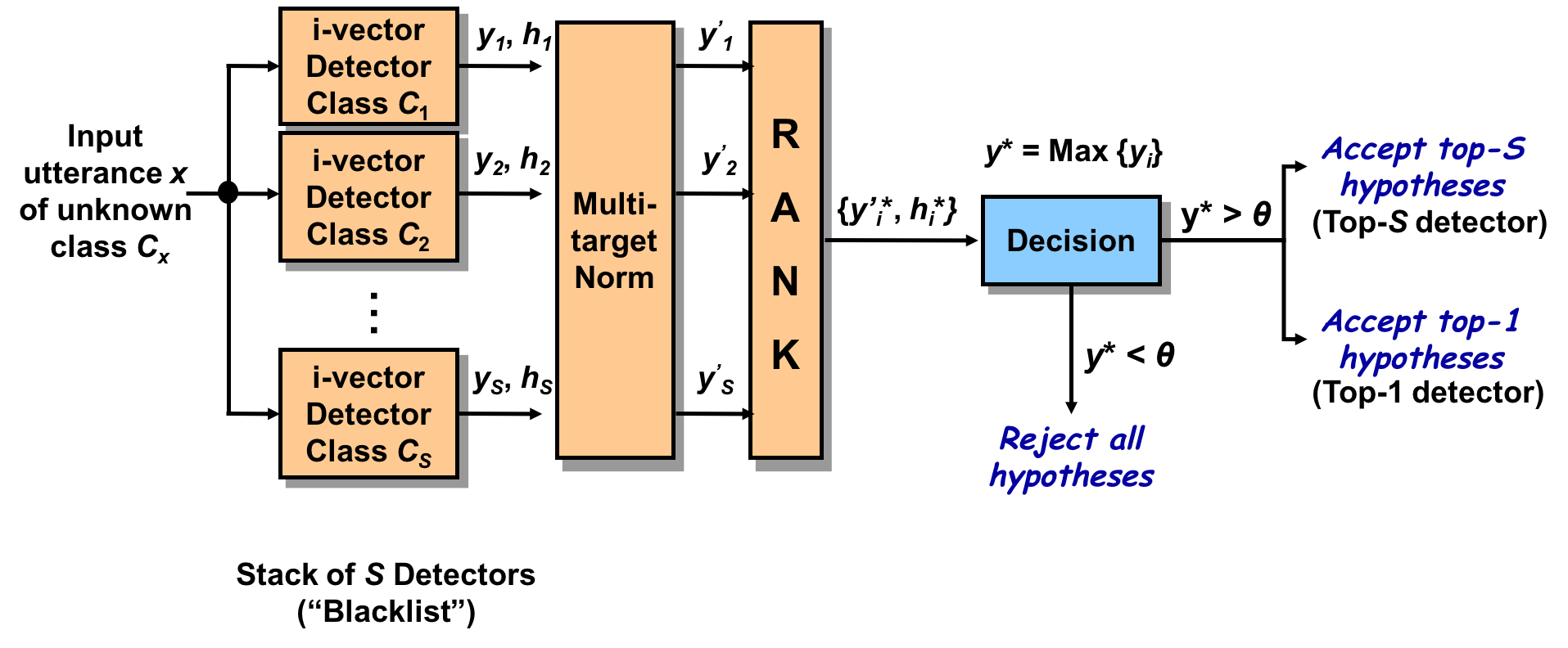}
    \caption{Multi-target Detector baseline for MCE 2018}
    \label{fig:baseline}
\end{figure}

The concept of baseline system is from the multi-target detector in \cite{Singer2004}. For each input, we rank the multi-target detector scores and accept the top-k hypotheses if the rank-1 score is above a detection threshold. If $k$ is the size of our blacklist ($S$), the system only cares if input is from anyone in the blacklist or not (top-$S$ detector). If $k$ is 1, the system further needs to determine who on the blacklist is speaking (top-$1$ detector). 

Additionally, multi-target score normalization (M-Norm) is applied to reduce the variability of decision score on multi-target. The purpose of M-Norm is shift and scale of score distribution between multi-target speakers and multi-target utterance to standard normal distribution. 

Suppose $x$ is input utterance of unknown class $C_x$ and multi-target (blacklist) speaker class set is $\{C_1,C_2,...,C_S\}$ where $S$ is size of multi-target speakers. $y_i$ is score of input x of detector class $C_i$ and can be represented as $y_i=score(C_i,x)$. The M-Norm score of $y_i$ is

\begin{equation}\label{eq:mnorm}
y_i' = score_M(C_i,x)=\frac{score(C_i,x)-\mu_M(i)}{\sigma_M(i)}
\end{equation}
The parameters of M-Norm are as follows:

\begin{equation}
\mu_M(i) = \frac{1}{||I||}\sum_{x\in\{C_1,...C_S\}} score(C_i,x) 
\end{equation}
\begin{equation}
\sigma_M(i) = \sqrt{\frac{1}{||I||} \sum_{x\in\{C_1,..,C_S\}} (score(C_i,x) - \mu_M(i))^2}
\end{equation}
where $||I||$ is a total number of utterances spoken by multi-target speakers. From the empirical experimental result, only shifting with $\mu_M$ or only scaling with $\sigma_M$ shows slightly better performance, but we provide baseline code with regular M-Norm equation \ref{eq:mnorm}.

\section{Submission of Results}
Evaluation results should be contained in ASCII file. Each line should contain 3 entries: [test set utterance ID],[score],[Closest blacklist speaker ID].\\

\noindent
\texttt{aacn\_382801,1.2345,01234567}\\
\texttt{zzow\_918095,0.6789,76543210}\\

The blacklist speaker ID is 8-digit number in \texttt{bl\_matching} file. For all test utterance, closest blacklist speaker ID should be specified even if the score is very low to enable computation of the Top-1 stack detector performance.
Participants can submit up to six file as described in section 2 and the file name should be \texttt{[team name]\_[training condition]\_[submission type]} format. 
For example,\\

\texttt{mit\_fixed\_primary}

\texttt{mit\_fixed\_contrastive1}

\texttt{mit\_fixed\_contrastive2}

\texttt{mit\_open\_primary}

\texttt{mit\_open\_contrastive1}

\texttt{mit\_open\_contrastive1}\\

\noindent
All participants also need to submit system description document and the name should be \texttt{[team name]\_description.pdf} format. 
For example,\\

\texttt{mit\_description.pdf}\\

\noindent
We recommend to use MCE2018 template which is based on the ICASSP 2019 regular format. The description should cover system's components including approaches, algorithms, configuration, parameters and data usage. Submission without a system description will not be counted in rankings and awards.

All files should be compressed in \texttt{.zip} or \texttt{.tar.gz} format. The single compressed file need to be submitted to MCE organizer by uploading on Dropbox folder. The Dropbox uploading link will be updated on the website.
%email : \texttt{\bf mce@lists.csail.mit.edu}. 

We have group email for general announcement and open discussion :
\texttt{\bf mce-2018@googlegroups.com}. All participants will be invided to this group shortly after registration. To ask any question to organizer, send to \texttt{\bf mce@lists.csail.mit.edu}\\

\section{Important dates}
% Case 1) If we can use on-line leaderboard, we should divide the test set into progress and evaluation set by 30/70 or 50/50.

%%%%
%\begin{itemize}
%\item Dataset release : Apr. 13
%\item Evaluation deadline(with system description) : May. 30
%\item Final evaluation announcement : Jun. 6
%\item (SLT Special Session/Tutorial Proposals deadline : Jun. 8)
%\item (SLT Special Session/Tutorial Proposals Notification : Jun. 22)
%\item Overview draft release : Jun. 22
%\item Paper submission on special session@SLT : Jul. 2
%\end{itemize}
%Case 2) If we cannot use on-line leaderboard, we should release test set before 1 or 2 weeks before evaluation deadline.

\begin{itemize}
\item May  21, 2018 : Development dataset release
\item Sep. 10, 2018 : Evaluation dataset release
\item Sep. 17, 2018 : Submission deadline for evaluation
\item Sep. 21, 2018 : Release evaluation dataset key
\item Oct. 10, 2018 : Result announcement (Rankings) 
\item Oct. 29, 2018 : (Recommend) Submit your findings to ICASSP 2019 Regular paper
\item Nov.  2018 : Awards presentation at the IEEE ICDM 2018 
\end{itemize}

\bibliographystyle{IEEEbib}
\bibliography{Odyssey2018_BibEntries}

% \section{Possible publicity place}
% \begin{itemize}
% \item SRE mailing list : sre16\_list@nist.gov
% \item SRE google group : \\sre16-participants@googlegroups.com
% \item ASVSpoof 2017 list : \\asvspoof-2017-list@googlegroups.com
% \end{itemize}

\end{document}